\documentclass{edp-conf}
\usepackage{graphicx}
\begin{document}

\TitreGlobal{SF2A 2003}

\title{Dust emission in the diffuse interstellar medium from far-infrared to 
millimeter: new constraints from WMAP}
\author{Lagache, G.}\address{IAS, B\^at 121, Universit\'e de Paris XI, 91405 ORSAY Cedex, France}
\runningtitle{New constraints from WMAP on the diffuse IS dust emission}
\setcounter{page}{237}
\index{Lagache, G.}

\maketitle
\begin{abstract}
This paper proposes a short review on our knowledge on the dust emission in the {\it diffuse} sky
from far-IR to the mm wavelength. 
The current understanding is based mainly on the analysis of the {\it COBE} data combined 
with several templates of gas tracers (HI and H${\rm \alpha}$). In the millimeter, 
emphasis is given to the new {\it WMAP} results on the so-called ``anomalous 
microwave emission''. 
\end{abstract}
%
\section{Introduction}
The dust emission from the far-InfraRed (far-IR) to the millimeter (mm) is intensively used
to trace the matter and the physical conditions in the Universe.
The objective of studying the dust emission at these wavelengths in 
the {\it diffuse} interstellar medium (opacities A$_{\rm v}$ lower than 
$\sim$0.3) of the Galaxy is twofold. First, 
unlike in the galactic plane where several clouds overlap on the lines
of sight, results are easier to interpret in term of physical properties 
of dust. Second, one of the major challenges in high sensitivity cosmic 
(microwave or submm) background anisotropy study is to determine 
the fraction of the observed signal due to diffuse galactic foregrounds.
\\
In the far-IR/submm, the emission comes from the large interstellar grains
emitting at the equilibrium temperature set by the balance between
heating and cooling. So far, only two experiments have measured the
large grain emission on large scale in the diffuse regions: IRAS
(only the 100 $\mu$m band is dominated by the large grain emission)
and {\it COBE} (mainly the {\it DIRBE} and {\it FIRAS} instruments).
In the mm, two experiments have covered the whole sky: {\it DMR/COBE} and
{\it WMAP}.
In this paper, we review the main results on the large grain far-IR
emission in the diffuse sky at high latitudes (Sect. 2).
In Sect. 3, we extend our understanding to longer wavelengths (submm and mm) 
and we finally discuss the new {\it WMAP} mm results in Sect 4.

\section{Far-IR dust emission associated with the HI gas}
At high latitudes, the determination of the galactic component relies on the existence
of a spatial correlation between gas and dust and thus of gas emission
lines with the associated dust emission. The correlation which has been the most
extensively investigated is that between IR emission and the 21 cm line from
atomic hydrogen. This correlation has been analysed for the whole dust spectrum using data
from the {\it DIRBE/COBE} and {\it FIRAS/COBE} experiments (Boulanger et al. \cite{Boul96}, Arendt et al. \cite{Arendt}).
The correlation is linear and very tight for HI column densities (N$_{\rm HI}$) lower than about
5 10$^{20}$ H/cm$^2$. For higher N$_{\rm HI}$, the data points depart
from the low emission correlation. 
Boulanger et al. (\cite{Boul96}) interpret this change in the slope 
as an increasing contribution of molecular gas for N$_{\rm HI}$ 
larger than 5 $10^{20}$ H cm$^{\rm -2}$. This interpretation is supported
by the work of Lagache et al. (\cite{Lagache98}): they showed that all regions 
with significant cold emission, that are associated with the molecular clouds, have an 
excess IR emission with respect to the high latitude far-IR/HI correlation.
In the low N$_{\rm HI}$ regions, the far-IR dust spectrum
is well fit by a single Planck curve with an emissivity proportional to $\nu^2$
and a temperature of 17.5 K. The dust emissivity per H atom is remarquably close to
the value obtained by Draine \& Lee (\cite{Draine84}) for a mixture of compact
graphite and silicate grains.\\
We have to note that the scatter in the far-IR/HI correlation at low
N$_{\rm HI}$ is larger than the noise. This can be due to several effects
as for example far-IR emission from ionised gas. Several searches for an IR emission
associated with the diffuse ionised gas has been prompted by the recent
H$_{\rm \alpha}$ surveys. Using the {\it WHAM} H$_{\rm \alpha}$ survey (Haffner et al. \cite{Haffner03}), 
Lagache et al. (\cite{Lagache00}) show that 
about 25$\%$ of the far-IR dust emission at high galactic latitudes
is un-correlated with the HI gas.
However, at first order, and for the following (Sect. 4), it is important to note that this
far-IR dust emission associated with the low N$_{\rm HI}$ gas has a stable HI-normalised 
spectrum, not changing with increasing
opacity (Lagache et al. \cite{Lagache99}).

\section{Towards the submillimeter and the millimeter wavelengths}
In the submm, using the {\it FIRAS} data, Finkbeiner et al. (\cite{Fink99}) showed that the one temperature
fit can be significantly improved by including a second emission component with a low temperature
(T$\sim$9 K) and an IR/visible emissivity ratio one order of magnitude larger than that of the warmer component.
It is unclear yet what is the physical origin of such a cold component and in particular if it represents
the emission from grains that are cold due to large submm emissivity.
This second component with T$\sim$9 K can be seen as a submm excess with respect to the T=17.5 K $\nu^2$ modified
black body. Such an excess has also been detected in the Archeops data (Bernard et al., in prep)
and could be attributed 
to a temperature dependence of the dust submm emissivity spectral index (M\'eny et al., in prep). 
\\

At longer wavelength (in the mm), there are in addition to the dust emission
two identified galactic components: synchrotron and free-free.
Synchrotron radiation dominates radio-frequency surveys but
the spectral index steepens with frequency
and exhibits spatial variations which are poorly known (e.g. Bennett et al. \cite{Bennett03}).
Free-free emission has a well-determined spectral behavior
and templates are now available thanks to the
{\it WHAM} (Haffner et al. \cite{Haffner03}) and the {\it SHASSA} (Gaustad et al. \cite{Gaustad01})
H$_{\rm \alpha}$ surveys.\\ 

Cross-correlations of mm data with far-IR maps have revealed
the existence of a microwave emission component with
spatial distribution traced by these maps. This component
has a spectral index suggestive of free-free
emission and so has been first interpreted as free-free emission
(Kogut et al. \cite{Kogut96}). However, Kogut (\cite{Kogut99}) showed in small parts
of the sky covered by H$_{\rm \alpha}$ data 
that the microwave emission were consistently brighter than free-free
emission. This is confirmed more recently 
by Banday et al. (\cite{Banday03}).
Thus, the correlated component cannot be due to free-free emission alone. 
Moreover, it is also well in excess 
and spectrally very different
from what is expected from thermal dust emission
and synchrotron radiation. Due to its mysterious nature,
this component has been called the ``anomalous microwave emission''.
Its identification, followed by its modelisation,
is still a major challenge both for galactic studies and cosmological
analysis.
Recent works suggest that this anomalous far-IR correlated component originates from spinning
dust grain emission (Draine \& Lazarian \cite{DL98a}, De Oliveira-Costa et al. \cite{DOC02}), tentatively detected at 5, 8 and 10 GHz by 
Finkbeiner et al. (\cite{Finkbeiner02}). Only extremely small
dust grains rotate sufficiently rapidly to produce a non negligible emission.
Draine \& Lazarian (\cite{DL98a}) estimate the size distribution of ultrasmall
grains and their rates of rotation and show that 
the electric dipole radiation could explain the anomalous component.

\section{Analysis of the WMAP millimeter data at high galactic latitude}
The previous detections/interpretations of the anomalous component have not been
confirmed by the first analysis of the {\it WMAP} mm data.
Very recently, Bennett et al. (\cite{Bennett03}) using 
{\it WMAP} data do not find any evidence for the anomalous microwave
emission that is limited to $<$5$\%$ of the 9.1 mm foreground
emission. 
Their foreground component model comprises
only free-free, synchrotron and thermal dust emission.
In their analysis, most of the emission of the anomalous component is
attributed to synchrotron radiation. 
Unlike in Bennett et al. (\cite{Bennett03}), an analysis of the galactic
contributions to the mm sky, based
on {\it WMAP} data combined with several
templates of dust emission and gas tracers,
do find evidence for a residual microwave emission, over free-free, synchrotron
and far-IR dust emission (Lagache \cite{Lagache03}).
This work focusses only on the high latitude regions.
Since the HI-correlated dust emission is the dominant component
at high galactic latitude at IR/far-IR/submm 
wavelengths{\footnote{Except in very low N$_{\rm HI}$ regions where the 
Cosmic IR Background becomes an important contribution}},
they compute the emission spectrum of the dust/free-free/synchrotron components associated with 
HI gas from low to large  N$_{\rm HI}$.
They find a significant residual {\it WMAP} emission over the free-free, synchrotron and
the dust contributions from 3.2 to 9.1 mm that
(1) exhibits a constant 
spectrum from 3.2 to 9.1 mm and (2) significantly decreases in amplitude 
when N$_{\rm HI}$ increases,
contrary to the HI-normalised far-IR emission which stays rather constant (see Sect. 2).
It is thus very likely that the residual {\it WMAP} emission is not
associated with the Large Grain dust component.
The decrease in amplitude with increasing opacity ressembles in fact to the decrease of the 
transiently heated dust grain emission observed in dense interstellar clouds.
This is supported by an observed decrease of the HI-normalised 60 $\rm \mu$m 
emission with N$_{\rm HI}$.
On the possible models of the  ``anomalous microwave emission'' linked 
to the small dust particles are the spinning dust  and the excess
mm emission of the small grains. The small grains
are transiently heated when
an ultraviolet photon is absorbed. The mean interval between
successive ultraviolet photons is longer than the cooling time
and thus, between 2 impacts, the temperature of the particles 
is very low. Such particles
could therefore emit significant emission in the mm
channels.\\
In conclusion, the so-called ``anomalous microwave emission''
seems to be linked to the small interstellar
dust grains. Due to the unknown properties
of the small particles, the models of possible mm emission
of these particles have large uncertainties.


\end{document}